\documentclass[runningheads]{llncs}
\usepackage[utf8]{inputenc}
\usepackage[edges]{forest}
\usepackage{textcomp}     
\usepackage{listings}     
\usepackage{syntax}               
\usepackage{tikz}
\usepackage{tikz-qtree}
\usepackage{xspace}
\usepackage{algorithm2e}
\usepackage{multirow}

\usepackage{url}
\usepackage{hyperref}
\usepackage[lofdepth,lotdepth,font={small}]{subfig}
\usepackage{csquotes}

\usepackage[font={small}]{caption} 

\usepackage[colorinlistoftodos]{todonotes}

\definecolor{codegreen}{rgb}{0,0.6,0}
\definecolor{codegray}{rgb}{0.5,0.5,0.5}
\definecolor{codepurple}{rgb}{0.58,0,0.82}
\definecolor{backcolour}{rgb}{0.9,0.9,0.95}
\definecolor{darkblue}{rgb}{0.2,0.2,0.4}
\definecolor{lightblue}{rgb}{0.5,0.7,0.9}
\definecolor{darkgreen}{rgb}{0.3,0.5,0.3}

\makeatletter
\DeclareRobustCommand\onedot{\futurelet\@let@token\@onedot}
\def\@onedot{\ifx\@let@token.\else.\null\fi\xspace}
\makeatother
\def\eg{\emph{e.g}\onedot}
\def\ie{\emph{i.e}\onedot}
\lstdefinestyle{ocaml}{
    basicstyle=\footnotesize\ttfamily,
    backgroundcolor=\color{backcolour},
    commentstyle=\color{darkblue},
    keywordstyle=\bfseries\color{red},
    numberstyle=\tiny\color{codegray},
    firstnumber=1,
    stringstyle=\color{codepurple},
    morestring=*[d]{"},
    captionpos=b,
    keepspaces=true,
    numbers=left,
    numbersep=5pt,
    showspaces=false,
    showstringspaces=false,
    showtabs=false,
    tabsize=2,
    identifierstyle=\idstyle,
    morekeywords={type, let, fun, if, then, else, match, with, function},
    moredelim = [s][\color{darkblue}]{[@}{]},
    moredelim = [s][\color{gray}]{(\*}{\*)}
}
\makeatletter
\newcommand*\idstyle{%
        \expandafter\id@style\the\lst@token\relax
}
\def\id@style#1#2\relax{%
        \ifcat#1\relax\else
                \ifnum`#1=\uccode`#1%
                        \bfseries\color{darkblue}
                \fi
        \fi
}
\makeatother

\newlength{\myl}

\newenvironment{indentgrammar}[1]{\setlength{\myl}{\widthof{#1}+2em}
  \grammarindent\the\myl
  \begin{grammar}}
  {\end{grammar}}

\newcommand{\litteral}[1]{\textcolor{blue!90!black!90!}{‘#1’}}
\newcommand{\commentary}[1]{\hspace*{\fill}\color{gray}#1\normalcolor}

\begin{document}


\title{Automatic Synthesis of Random Generators for Numerically  Constrained Algebraic Recursive Types%
\thanks{this research was partially supported by the ANR PPS project ANR-19-CE48-0014 and the “DYNNET” project, co-funded by the Normandy County Council and the European Union (ERDF-ESF 2014-2020).}}

\author{%
  Ghiles Ziat\inst{1} \and
  Vincent Botbol\inst{2} \and
  Matthieu Dien\inst{3} \and
  Arnaud Gotlieb\inst{4} \and
  Martin P\'epin\inst{1} \and
  Catherine Dubois\inst{5}}

\institute{%
  Université Paris Cité, IRIF, 75205 Paris Cedex 13, France, \email{\{ghiles.ziat, martin.pepin\}@irif.fr},
  \and
  Nomadic Labs, \email{vincent.botbol@nomadic-labs.com}
  \and
  Normandie Université, UNICAEN, ENSICAEN, CNRS, GREYC, 14000 Caen, France, \email{matthieu.dien@unicaen.fr}
  \and
  SIMULA RESEARCH LAB, \email{arnaud@simula.no}
  \and
  Ecole Nationale Supérieure d'Informatique pour l'Industrie et l'Entreprise, Samovar, \email{catherine.dubois@ensiie.fr}
}

\maketitle{}

\begin{abstract}
In program verification, constraint-based random testing is a powerful
technique which aims at generating random test cases that satisfy
functional properties of a program. However, on recursive constrained
data-structures (e.g., sorted lists, binary search trees, quadtrees),
and, more generally, when the structures are highly constrained,
generating uniformly distributed inputs is difficult. In this paper,
we present Testify: a framework in which users can define algebraic
data-types decorated with high-level constraints. These constraints
are interpreted as membership predicates that restrict the set of
inhabitants of the type. From these definitions, Testify automatically
synthesises a partial specification of the program so that no function
produces a value that violates the constraints (\eg a binary search
tree where nodes are improperly inserted). Our framework augments the
original program with tests that check such properties. To achieve
that, we automatically produce uniform random samplers that generate
values which satisfy the constraints, and verifies the validity of the
outputs of the tested functions. By generating the shape of a
recursive data-structure using {\it Boltzmann sampling} and generating
evenly distributed finite domain variable values using constraint
solving, our framework guarantees size-constrained uniform sampling of
test cases. We provide use-cases of our framework on several key data
structures that are of practical relevance for developers. Experiments
show encouraging results.
\end{abstract}


\section{Introduction}
Software Testing is one of the most widespread program
verification techniques, and is also one of the most practical to
implement.  One interesting instance of it is
Property-based Testing (PBT), where programs are tested by
generating random inputs and results of the output are compared
against software specifications.  This technique has been
extensively studied, for testing correctness~\cite{Go05,Pa&al11},
exhaustiveness~\cite{Co08}, complexity~\cite{Ja09} etc. However,
this technique requires the developer to manually write the
tests, that is the properties to be checked and the random
generators. The latters can be particularly complicated to
design, especially in the case of complex and constrained
algebraic data structures.

In this field, {\it constraint-based random testing} \cite{Hughes07}
(commonly used in PBT \cite{Laplante07,Carlier10})
is a powerful technique which aims at generating random test cases
that satisfy functional properties of a program under test. By
specifying a property that a program has to satisfy and by using
uniformly-distributed inputs generators, it is possible to uncover
subtle robustness faults that may be not be discovered otherwise. For
instance, \cite{Abdellatif03} explored the usage of PBT for testing a 
steam boiler, \cite{Loscher17} explored its
usage for wireless sensor network applications. It is worth noticing
that generating inputs according to a uniform probability distribution
is crucial to ensure that all the distinct program behaviours have the
same chance to be triggered, even those which are the most
constrained. The technique has been successfully applied in the field
of unit testing for imperative programs \cite{Gotlieb10} as well as
various programming languages including Haskell~\cite{qcheck},
Prolog~\cite{Amaral14} and proof-assistant methodologies and tools
such as Coq~\cite{Paraskevopoulou15} or Isabelle~\cite{Bulwahn12}.
Sampling constraint systems solutions according to a uniform
distribution is a well-known difficult problem. Initially studied in
the context of hardware testing \cite{Naveh07}, the problem has been
studied in \cite{Gogate06} and more extensively in \cite{Allen17}.
Other random generation schemes are either not uniform, or very slow
\eg rejection sampling is generally uniform by construction, but fits
very poorly with generation under constraints.

Recently, in \cite{ZDB21} the authors introduced an automated
framework capable of providing tests for functions that manipulate
constrained values without requiring manual input from the
programmer. The framework introduces a type language, with algebraic
data-types, and constrained types \ie types augmented with a
membership predicate that is used to filter invalid
representations. To verify that a function does not create invalid
representations, the authors opted for a random testing approach. The
main interest of the framework is that both the generators and the
specifications are automatically extracted from the constraints
specified by the user, which greatly alleviate the user's
workload. Generators are uniform random value samplers used to provide
input for functions, and specifications that are predicates that
verify that a given value satisfies the constraint attached to its
type, are used to check whether a function's output violates the
constraint or not. Their tool is implemented as a pre-processor for
OCaml programs, \ie before compiling, programs are rewritten into
augmented programs where a test suite has been added. During the
pre-processing step, from each constrained type declaration $\tau$ is
extracted a CSP $p$. Then, each $p$ is solved only once, that is to
say that a characterisation of the set of solutions of $p$, called
coverage, is calculated. Each coverage is then compiled into code
which uniformly generates solutions which are then converted back into
values of the type $\tau$.  However, to be able to solve a CSP only
once per constrained type, the authors limit themselves to types
involving a fixed number of numerical atoms (\eg tuples), which
automatically excludes recursive types. This makes it impractical as,
for instance, in OCaml, real-world programs rely heavily on recursive
data-types (lists, trees, sets, etc.).

This paper investigates the automatic synthesis of uniform
pseudo-random generators, as in \cite{ZDB21}, but for recursive
constrained types.

\subsection{Contributions}
\begin{itemize}
\item A programmable method to restrict the values a recursive type
  can take.

\item An algorithm that uses Boltzmann generation and constraint
solving to automatically derive uniform generators for recursive
constrained types.

\item An experimental study of the performances of our technique.
\end{itemize}


\subsection{Outline} This paper is organised as follows:
Sec.\ref{sec:declprog} presents use cases of our methods on
some examples of realistic code.  Sec.\ref{sec:solving}
defines our solving technique which mixes Boltzmann generation
and global constraint solving.  Sec.\ref{sec:boltzmann}
recalls some elementary notions about Boltzmann sampling and
details some specifics about our use-case.
Sec.\ref{sec:implem} presents our prototype and gives some
details about its functioning, current capabilities and
restrictions. We also give some details about our implementation
and measure experimentally the performances of the generators we
derive for recursive constrained types.
Sec.\ref{sec:related} describes some related work. Finally,
Sec.\ref{sec:conclusion} summarises our work and discusses
possible future improvements.

\section{A Declarative Programming Approach}\label{sec:declprog}
We propose a testing framework that allows programmers to specify
constraints on recursive data structures. From these constraints, the
framework extracts a Constraint Satisfaction Problem (CSP) which is
solved in such a way that uniform random instances (i.e., test cases)
are generated. These instances are then used for testing functions in
order to find defects.

\subsection{Preliminaries}
A pseudo-random generator $g$ for an algebraic data-type $\tau$ is a
function $g$ of type $\mathcal{S} \rightarrow \tau$.  Here,
$\mathcal{S}$ is the random state used by the pseudo random number
generator. A constrained type is a pair $\langle \tau , p \rangle$,
with $\tau$ an algebraic data-type and $p : \tau \rightarrow bool$ a
predicate over values of type $\tau$.  The set of its inhabitants is
defined as $\{t \in \tau~|~p(t) = true\}$.  A pseudo-random generator
$g$ for a constrained type $\langle \tau , p \rangle$ is a function $g
: \mathcal{S} \rightarrow \tau$ s.t $\forall s \in \mathcal{S},
p(g(s)) = true $.

Here, we face two main challenges for automating random testing of
recursive data-types. First, we have to equip the developer with
convenient means for specifying constraints attached to a given
data-type. For example, we want to express that a list of integers is
sorted or that a tree is a binary search tree (i.e., the left child
node value is always smaller than the right one). Second, building an
uniform random value generator for constrained recursive data-types is
highly challenging. Recursive types can dynamically grow to an
arbitrarily large size and, deriving generators for such types
requires the resolution of a complex constraint system. In particular,
we have to manage CSPs with an a-priori unknown number of variables
and constraints.
The grammar of Ocaml types and constraints annotations are given in Figure~\ref{fig:adt}. In the following, we give two illustrative examples.
\subsection{Example 1: Inserting an element into a set of integers}
Let start with {\tt list}, a recursive data-type associated to lists
of integers, for which a possible type declaration is given in
Fig.\ref{ilists}.
\begin{figure}[h]
\begin{lstlisting}{ocaml}
type list = Empty | Cons of int * list
\end{lstlisting}\caption{OCaml type declaration of lists of integers\label{ilists}}
\end{figure}
Using {\tt list} to specify a Set data structure can easily be done
using Testify, by using the annotation \lstinline{[@@satisfying _]}
and the \lstinline{[alldiff]} constraint as illustrated by
Fig.~\ref{uilists}.
\begin{figure}[h]
\begin{lstlisting}{ocaml}
type uniquelist =
  | Empty
  | Cons of int * uniquelist [@@satisfying alldiff]
\end{lstlisting}
\caption{OCaml type declaration of sets of integers using lists\label{uilists}}
\end{figure}

We can automatically test the functions that manipulate instances of
the \lstinline{uniquelist} type by checking if they break the
properties attached to it. For that, we have to define a generator and
a specification for the corresponding type. To randomly generate
instances, we first draw at random an instance of size $n$ using
Boltzmann generation (see Sec.\ref{sec:boltzmann}), then we build a CSP
$\mathcal{(X,D,C)}$ containing $n$ finite domain variables and solve
it using Path-oriented Random Testing (PRT) (see
Sec.\ref{sec:solving}) and eventually we build a random generator $g$
able to produce uniformly distributed sets of size $n$.

A key aspect of Testify is based on the usage of {\it global
  constraints}, which are arithmetic-logic constraints holding over a
non-fixed number of variables. In the example of Fig.\ref{uilists}, we
translate the declaration \lstinline{[alldiff]} into a {\tt
  all_different} global constraint implementation and used it to
generate uniformly distributed solutions that can be used to polulate
test cases. For other recursive data-types, we use combination of
multiple global constraints and arithmetic constraints.  Possible
recursive data-types that can be implemented and tested in our
framework include functions that generate and manipulate (un-)ordered
lists and sets, trees, binary search trees, quadtrees, etc.

Fig.\ref{addlist} shows an example of a function which implements the
insertion of an element within a set of integers and the code that is
automatically generated for the testing of this function\footnote{The
predicate {\tt alldiff_checker} checks that the result list does not
contain duplicates. It should not be mixed with the version of {\tt
  alldiff} used in the type declaration which is used to generate
randomly distributed solutions of that constraint}.

\begin{figure}[ht]
\begin{lstlisting}{ocaml}
let rec add (x:int) (l:uniquelist) : uniquelist =
  match l with
  | Empty -> Cons(x,Empty)
  | Cons(h,tl) -> if x <> h then Cons(h,(add x tl) else l)

(* generated code*)  
let add_test () =
  let size = Random.int () in let rand_x = Random.int () in 
  let rand_l = unique_list size in
  assert (alldiff_checker (add rand_x rand_l))  
\end{lstlisting}%
\caption{Insertion of an element into a set, and the generated
  corresponding test\label{addlist}}%
\end{figure}

Here, testing the function function means verifying that every output produced is indeed sorted 
(\lstinline{assert (alldiff_checker (add rand_x rand_l))}. 
Note that we have used the return type annotation to automatically derive a (partial) specification for 
the function, but the generator we automatically synthesise can also be used to
test any hand-written specification.

\subsection{Example 2: Binary Search trees}
Binary Search Trees (BST) are binary trees that additionally satisfy
the following constraint: the key in each node is greater than or equal to any
key stored in the left sub-tree, and less than or equal to any key
stored in the right sub-tree. Stated differently, the keys in the tree
must be in increasing order in a depth-first search traversal, in
infix order. From this observation, we propose, using our
framework, a possible OCaml declaration for BSTs
illustrated in Fig.\ref{bstrees}.

\begin{figure}[ht]
\begin{lstlisting}{ocaml}
type bst =
  | Node of bst * (int[@collect]) * bst
  | Leaf [@@satisfying fun x -> increasing x)]
\end{lstlisting}
\caption{Testify type annotation for binary search trees\label{bstrees}}
\end{figure}

Rather than defining global constraint for all user-declared
data-types, we break the problem in two parts. On the one hand,
we define or reuse known global constraints for lists, and on the
other hand we define a way to browse data structures, in a certain
order, by collecting the components that are subject to a global
constraint. This is done using the \lstinline{(int[@collect])}
annotation.

Also, the order in which the structure is explored is crucial as it
determines the order in which the variables will be passed to the
global constraint. By default, a depth first order is assumed. For
constructors with several arguments (\eg \lstinline{Node}), and for
tuples, the order in which the traversal is made is mapped on the
declaration order of the tuple component, that is in traversal
order. Fig.\ref{collector} shows the code generated that
traverses the tree.

\begin{figure}[ht]
\begin{lstlisting}{ocaml}
let rec collect = function
| Node (a, b, c) ->
    List.flatten [collect a; Collect.int b; collect c]
| Leaf -> []
\end{lstlisting}
\caption{Generated collector for binary trees\label{collector}}
\end{figure}

Here, the primitive \lstinline{Collect.int} is a primitive of our
framework that takes an integer and builds the singleton list with
this element.  This way, we first visit the left sub-tree, the root
and the right sub-tree. Using pre-order or post-order would give
different results. This means that the constructor \lstinline{Node}
must be declared in the above order and, for example, the following
would be invalid: \lstinline{Node of (int[@collect]) * binary_tree * binary_tree}.
However, this restriction can easily be lifted by providing an
annotation which would allow the programmer to explicitly specify the
traversal order. Similarly, a global annotation \lstinline{[@bfs]}
(resp. \lstinline{[@dfs]}) could be used to specify that the structure
must be traversed using a breadth first search (resp. depth first
search). This will be studied in future work.

\section{Constrained Type Solving}\label{sec:solving}
A {\it Constraint Satisfaction Problem} (CSP) is a triple $\mathcal{(X,D,C)}$ where $\mathcal{X}$ is a set of variables, $\mathcal{D}$ is a function associating a finite domain (considered here as a subset of $\mathcal{Z}$ without any loss of generality) to every variable and $\mathcal{C}$ is a set of constraints, each of them being $<var(c),rel(c)>$, where $var(c)$ is a tuple of variables $(X_{i_1},..,X_{i_r})$ called the scope of $c$, and $rel(c)$ is a relation between these variables, i.e., $rel(c) \subseteq \prod_{k=1}^r D(X_{i_k})$. For each constraint $c$, the tuples of $rel(c)$ indicate the allowed combinations of value assignments for the variables in $var(c)$. Given a CSP $\mathcal{(X,D,C)}$, an {\it assignment} is a vector $(d_1,..,d_n)$, which associates to each variable $X_i \in \mathcal{X}$ a corresponding domain value $d_i \in D(X_i)$. An assignment satisfies a constraint $c$ if the projection of $\mathcal{X}$ onto $var(c)$ is a member of $rel(c)$. The set of all satisfying assignments is called the solution set, noted $sol(\mathcal{C})$.
A constraint $c$ is said to be {\it satisfiable} if it contains at least one satisfying assignment, it is inconsistent otherwise. A CSP $\mathcal{(X,D,C)}$ is satisfiable if it contains at least one assignment which satisfies all its constraints (i.e., $sol(\mathcal{C}) \neq \emptyset)$. A global constraint is an extension of CSPs with relations concerning a non-fixed number of variables. For instance, the {\it sort(Xs, Ys)} global constraint \cite{Mehlhorn00} takes as inputs two lists of $n$ finite domain variables $Xs, Ys$ (where $n$ is unknown) and states that for each satisfying assignment $(d_1,..,d_n, d'_1,..,d'_n)$ of the constraint, we have $\forall j, \exists i$ s.t. $d'_j=\sigma(d_i)$ and $d'_1 \leq .. \leq d'_n$, where $\sigma$ is a permutation of $[1..n]$. Filtering a global constraint $c(X_1,..,X_n)$ with the {\it bound-consistency} local filtering property means to find $D'$ such that for all $i$, 
the extrema values of $D'(X_i)$ are parts of satisfying assignments of $c$.  
 
\subsection{Path-Oriented Random Testing}
{\it Path-oriented Random Testing} (PRT) is basically a
divide-and-conquer algorithm, introduced in \cite{Gotlieb10}, which
aims to generate a uniformly distributed subset of solutions of a
CSP. Starting from an initial filtering step result, the general idea
is to fairly divide the resulting search space into boxes of equal
volumes and, after having discarded inconsistent boxes using
constraint refutation, to draw at random satisfying assignments.

More precisely, applying constraint filtering results in domains that
can be over-approximated by a larger box (i.e., an hyper-cuboid) that
contains all the filtered domains. Based on an external division
parameter $k$, PRT then fairly divides the box into $k$ subdomains of
\emph{equal} volume.  When a subdomain cannot be divided according to
the division parameter $k$, then it is simply extended until its area
can be divided. The iteration of the process leads to a fair partition
of the search space into $k^n$ subdomains where $n$ is the number of
variables of the CSP. Then constraint refutation can be used to
discard (some) subdomains which are inconsistent with the rest of the
CSP. As all subdomains have the same volume, it becomes possible to
sample first the remaining subdomains and then, second, to randomly
draw values from these subdomains. Note that, when all the subdomains
are shown to be inconsistent, then the CSP is shown to be
inconsistent. This contrasts with reject-based methods which will
trigger assignment candidates and will reject them afterwards, without
terminating in a reasonable amount of time.

\begin{algorithm}[ht]
\caption{Path-Oriented Random Testing adapted from \cite{Gotlieb10} to the uniform random generation of $N$ solutions of a CSP}\label{PRTalgo}
\SetKwInOut{Input}{Input}
\Input{CSP:$\mathcal{(X,D,C)}, k, N$: \#Sol. - {\bf Output:} $t_1,..,t_N$ or $\emptyset$ (Inconsistent)}
        $\mathcal{D'} := {\tt boxfilter_{bc}}\mathcal{(X,D,C)}$; 
        $(H_1,..,H_{p}) := {\tt Fairly\_Divide}(\mathcal{D'},k)$; 
        $T := \emptyset$;\\
         \While{$N>0$ and $p \neq 0$}{
            Pick up uniformly $H$ at random from $H_1,..,H_p$;\\
            \eIf{$H$ is inconsistent w.r.t. $\mathcal{C}$} 
                {remove $H$ from $H_1,..,H_p$}
                {Pick up uniformly $t$ at random from $H$ and remove it;\\
                  \If{$\mathcal{C}$ is satisfied by $t$} 
                      {add $t$ to $T$ ; $N$ := $N-1$;}
                }
            }
        \Return T;
\end{algorithm}

The PRT algorithm, adapted from \cite{Gotlieb10} to the case of CSP
solution sampling, is given in Figure~\ref{PRTalgo}. It takes as
inputs a CSP, a division parameter $k$, and $N$ a non-negative
integer. Here, we make the hypothesis that, if the CSP is consistent,
it contains more than $N$ solutions.  The algorithm outputs a sequence
of $N$ uniformly distributed random assignments which satisfy the
CSP. If the CSP is unsatisfiable, then PRT returns $\emptyset$.  After
an initial filtering step using bound-consistency, the algorithm
partitions the resulting surrounding box in subdomains of equal volume
({\tt Fairly\_Divide} function). Then, for each locally consistent
subdomain $H$ in the partition, value assignments are randomly
selected and checked against the constraints of the CSP.  Those which
do not satisfy the constraints are simply rejected. As shown in
\cite{Gotlieb10}, this process ensures the uniform generation of
tuples in the solution space. 

\subsection{Extension with Global Constraints}

Handling global constraints is a natural extension of PRT as it allows
us to handle recursive constrained data-types. As the shape of the
data structure is unknown at constraint generation time, the number of
variables to be handled is also unknown in the general case. Thus,
using global constraints in this context is particularly useful as it
allows us to avoid the decomposition of a global constraint into 
the conjunction of several simpler constraints. This results in both a 
stronger and faster pruning. In
order to handle recursive constrained data-types, we had to provide a
dedicated interface for accessing the deductions from global
constraint solving. To facilitate the access to global constraints, we
created an API which provides results of PRT over different global
constraint combinations. The API provides access to predicates such as
{\tt increasing_list(+int LEN,+int GRAIN,-var L)} in which {\tt L} is
instantiated to a list of {\tt LEN} uniformly distributed random
integers ranked in increasing order, and the random generator is
initialised with {\tt GRAIN}. Optionally, the predicate can be called
with domain constraints in order to constrain the returned list of
values in specific subdomains. Other similar predicates are provided
as part of the API, namely {\tt increasing_strict_list/3 (+int
  LEN,+int GRAIN,-var L)} which returns a list of strictly increasing
integers; {\tt decreasing_list/3} (resp. {\tt
  decreasing_strict_list/3}) which provides a list of integers in
(resp. strict) decreasing order or else {\tt alldiff_list/3} which
returns a list of uniformly distributed random distinct integers. PRT
can also be used in combination with any available global constraint
and arithmetico-logic constraint. The following example, given in
Fig.\ref{fig:PRTGlob} illustrates how PRT is used in this respect.

\begin{figure}[ht]
  \centering
  \subfloat[\centering Shape generation]{
    \begin{tikzpicture}[scale=0.8,level distance=1cm,
        every node/.style={minimum size=5mm,fill=blue!10,circle,inner sep=1pt},
        level 1/.style={sibling distance=1.6cm},
        level 2/.style={sibling distance=1cm},
        edge from parent path={(\tikzparentnode) to (\tikzchildnode)}
      ]
      \node(x4){?}
      child {node(x2){?}
        child {node(x1){?}}
        child {node(x3){?}}
      }
      child {node (x6){?}
        child {node (x5){?}}
        child[missing]
      };
      
    \end{tikzpicture}
  }
  \qquad
  \centering
  \subfloat[Depth-first walk]{
    \begin{tikzpicture}[scale=0.8,level distance=1cm,
        every node/.style={minimum size=5mm,fill=blue!10,circle,inner sep=1pt},
        level 1/.style={sibling distance=1.6cm},
        level 2/.style={sibling distance=1cm},
        edge from parent path={(\tikzparentnode) to (\tikzchildnode)}
      ]
      \node(x4){\small $Y_4$}
      child {node(x2){\small $Y_2$}
        child {node(x1){\small $Y_1$}}
        child {node(x3){\small $Y_3$}}
      }
      child {node (x6){\small $Y_6$}
        child {node (x5){\small $Y_5$}}
        child[missing]
      };
      \path[->,thin,gray](x1) [out=30, in=-100] edge (x2); 
      \path[->,thin,gray](x2) [out=-80, in=140] edge (x3);
      \path[->,thin,gray](x3) [out=95, in=-95] edge (x4);
      \path[->,thin,gray](x4) [out=-85, in=95] edge (x5);
      \path[->,thin,gray](x5) [out=40, in=-90] edge (x6);
    \end{tikzpicture}
  }
  \qquad  
  \centering
  \subfloat[Generating the corresponding CSP]{%
  \small
\begin{tabular}{c}
$X_1 \in -2..8,X_2 \in -3..5,X_3 \in -3..10$,\\
$X_4 \in -1..9, X_5 \in 0..7,X_6 \in 0..8$,\\
sort($(X_1,..,X_6),(Y_1,..,Y_6)$),\\
$Y_2 = Y_1+Y_3, Y_5 = Y_6, Y_4 = Y_2 + Y_6$ 
\end{tabular}
  }

  \qquad
  \centering
  \subfloat[PRT first sol.]{
    \begin{tikzpicture}[scale=0.8,level distance=1cm,
        every node/.style={minimum size=5mm,fill=blue!10,circle,inner sep=1pt},
        level 1/.style={sibling distance=1.6cm},
        level 2/.style={sibling distance=1cm},
        edge from parent path={(\tikzparentnode) to (\tikzchildnode)}
      ]
      \node(x4){\small $8$}
      child {node(x2){\small $-1$}
        child {node(x1){\small $-1$}}
        child {node(x3){\small $0$}}
      }
      child {node (x6){\small $9$}
        child {node (x5){\small $9$}}
        child[missing]
      };
    \end{tikzpicture}
    }
    \qquad
  \centering
  \subfloat[PRT second sol.]{
    \begin{tikzpicture}[scale=0.8,level distance=1cm,
        every node/.style={minimum size=5mm,fill=blue!10,circle,inner sep=1pt},
        level 1/.style={sibling distance=1.6cm},
        level 2/.style={sibling distance=1cm},
        edge from parent path={(\tikzparentnode) to (\tikzchildnode)}
      ]
      \node(x4){\small $4$}
      child {node(x2){\small $0$}
        child {node(x1){\small $0$}}
        child {node(x3){\small $0$}}
      }
      child {node (x6){\small $4$}
        child {node (x5){\small $4$}}
        child[missing]
      };
    \end{tikzpicture}
    }
    \qquad
  \centering
  \subfloat[PRT third sol.]{
    \begin{tikzpicture}[scale=0.8,level distance=1cm,
        every node/.style={minimum size=5mm,fill=blue!10,circle,inner sep=1pt},
        level 1/.style={sibling distance=1.6cm},
        level 2/.style={sibling distance=1cm},
        edge from parent path={(\tikzparentnode) to (\tikzchildnode)}
      ]
      \node(x4){\small $4$}
      child {node(x2){\small $-1$}
        child {node(x1){\small $-2$}}
        child {node(x3){\small $1$}}
      }
      child {node (x6){\small $5$}
        child {node (x5){\small $5$}}
        child[missing]
      };
    \end{tikzpicture}
    
  }
  \caption{Generation of a constrained BST of size 6. Division
    parameter=2, length of seq.=3, 60 subdomains over 64 have been
    discarded after the first filtering.}\label{fig:PRTGlob}
\end{figure}
In this example, PRT is used with one global constraint, namely
$sort(Xs,Ys)$, and some domain and arithmetic constraints to populate
a constrained binary search tree (BST) of size~$6$. In this example,
the shape of the BST is unknown and some constraints hold over the
keys: the domain of the key-variables is constrained (from an
externally specified source), \eg{}, key $X_1 \in -2..8$, key $X_2 \in
-3..5$, etc. and any key of the BST corresponds to the sum of its
children (if any), \eg{}, $Y_{father} = Y_{child_l} + Y_{child_r}$. Note that
the keys have to be set in increasing order to correspond to a valid
BST. Note also that we ignore in which order will the keys be
positioned in the tree. The first step of our method corresponds to
the generation of a uniformly distributed random shape of the BST
(Fig.\ref{fig:PRTGlob}(a)) using the Boltzmann method, described in
Sec.\ref{sec:boltzmann}. Then, a depth-first walk along the tree
assigns variable identifiers to the nodes and collects the constraints
that must hold over the constrained data structure
(Fig.\ref{fig:PRTGlob}(b)). The generated CSP
(Fig.\ref{fig:PRTGlob}(c)) can then be solved by using PRT, which
generates, in this example, three uniformly distributed random
solutions (Fig.\ref{fig:PRTGlob}(d)(e)(f)). It is worth noticing that
other uniform random solutions sampling methods such as
\cite{Gogate06,Vavrille21} could have been used in this context. PRT
was chosen because of its availability and simplicity. However,
non-uniform random sampling such as a simple heuristic selecting at
random variable and values to be enumerated first would not have been
appropriate in this context as the goal was to test the robustness of
user-defined functions in functional programming.

\section{Boltzmann Sampling}\label{sec:boltzmann}
The Boltzmann method was introduced in~\cite{DuFlLoSc04} as an
algorithmic method to derive efficient sampler from
\emph{combinatorial classes}. Combinatorial classes are just sets of
discrete structures with a size (a non-negative integer) and such that
the number of structures having the same size is finite.  For example,
the binary trees whose size is the number of leafs is a combinatorial
class, but binary trees whose size is the length of leftmost branch is
not because the number of binary trees with a leftmost branch of fixed
length $k$ is infinite.  We briefly present the method here and refers
the reader to~\cite{DuFlLoSc04} for more details.
\begin{figure}[htp]
  \begin{indentgrammar}{type}
    <decl> ::= \litteral{type} <type identifier> \litteral{=} <type> \commentary{type declaration}\\
    \{\litteral{[@@satisfying} <constraints> \litteral{]}\} \commentary{Testify's annotation}

    <type> ::= <coretype>
    \alt <sumtype> 
    
    <coretype> ::=   \litteral{int} | \litteral{float} | \litteral{char} | \ldots \commentary{basic types}
    \alt <coretype> \{\litteral{*} <coretype> \} \commentary {product}
    \alt <type identifier>
    
    <sumtype> ::= <constructor identifier> \{\litteral{[@collect]}\} \litteral{of} <coretype>
    \alt <sumtype> \{\litteral{|} <sumtype> \}
    
    <constraints> ::= \litteral{alldiff} | \litteral{increasing} | \litteral{decreasing} \commentary{SICStus global constraints}
    \alt <arith> \commentary{arithmetic constraints like in \cite[Fig. 1]{ZDB21}}
    \alt <constraints> \{\litteral{\&\&} <constraints>\}
  \end{indentgrammar}
  \caption{Syntax of OCaml algebraic data-types (ADT)\label{fig:adt} with Testify's annotation}
\end{figure}

In the context of that paper (similarly to~\cite{CaDa09}), that method
directly translates into an automatic way to derive a uniform random
generator of terms for the type language whose syntax is given
in Fig.\ref{fig:adt}. In our case, the produced generators only generate a
\emph{shape} of tree structure in a first step and the content of this
shape is provided in a second step by a constraint solver which makes
sure to fill the shape with values that satisfy the specified
constraints. For each constrained recursive type declaration, we must
therefore generate a glue function between the shapes generated by the
Boltzmann sampling method and the solutions returned by the solver
used. This function is illustrated in the case of binary search trees
in Fig.\ref{fig:shapefill}

\begin{figure}
\begin{lstlisting}{ocaml}
  let rec fill_binary_tree shape solutions =
    match shape with
    | Label ("Node", [x1; x2; x3]) ->
        let x1 = fill_binary_tree x1 solutions in
        let x2 = Testify_runtime.to_int x2 solutions in
        let x3 = fill_binary_tree x3 solutions in
        Node (x1, x2, x3)
    | Label ("Leaf", []) -> Leaf)
\end{lstlisting}
\caption{Generated function for filling the shapes for binary search trees.\label{fig:shapefill}}
\end{figure}

Here, we consider types as sets of terms (the inhabitants of the type)
whose size is the number of \verb|[@collect]| values they
contain. For example, using \lstinline|type binary_tree| of
Sec.\ref{sec:declprog}, the term
\lstinline|Node(Node(Leaf, 3, Leaf), 25, Leaf)|
has size $2$.

In the following, we denote $\Gamma\mathcal{A}_x$ a Boltzmann sampler
of parameter $x$ for the set $\mathcal{A}$. Such sampler produces an
object $\gamma \in \mathcal{A}$ with a probability
$\frac{x^{|\gamma|}}{A(x)}$ where $|\gamma|$ is the size of $\gamma$
and $A(x)$ is a normalizing factor called \emph{generating
  series}\footnote{The generating series $A(z)$ of a combinatorial
  class $\mathcal{A}$ is defined by
  $A(z) = \sum_{\gamma \in \mathcal{A}} z^{|\gamma|}$}.
Note that objects of the same size have the same probability to be
drawn.

The second interest of Boltzmann samplers is that they compose well
with sum, product and substitution \emph{i.e.} the constructors of  ADTs.
Fig.\ref{boltzsampl} shows the derivation of such samplers.
\begin{figure}[htb]
\begin{lstlisting}[mathescape=true]{ocaml}
type t = a * b (* a and b are 2 types previously defined *)
let gen_t x = gen_a x, gen_b x

type u = A of a | B of b
let gen_u x = 
  if random() < ${A(x)}/{(A(x) + B(x))}$
  then gen_a x else gen_b x

type alst = Nil | Cons of a * alst (* $alst(z) = 1 + z \cdot alst(z)$ *)

let rec gen_alst x : alst =
  if random() < ${1}/{(1 - A(x))}$ 
  then Nil else Cons(gen_a x, gen_aList A(x))
\end{lstlisting}\caption{Sampler derivation using Boltzmann\label{boltzsampl}}\end{figure}
At the end of the generating process, the object drawn has a random
size, but we see in the previous code that the choice of the
parameter $x$ influences the size.
Note that we can precisely and efficiently
compute $x$ to target a size (see \cite{BeBoDo18} or \cite{PiSaSo12} for the details).

Still, the size is random. The last ingredient is to choose a
parameter $\epsilon$ (which does not depend of the targeted size $n$)
and keep only objects of size between $n-\epsilon$ and $n+epsilon$.
Thus, the size of the object is kept up to date during the generation
and the generation is stopped if that size exceeds the upper bound
$n+\epsilon$. At the end, the object may be smaller than $n-\epsilon$
in which case it is rejected too.
However, the theory (see~\cite{DuFlLoSc04}) guarantees that the
rejections cost remain relatively low, \emph{i.e.} the cumulated size of objects
sampled to obtain an object of size in the interval
$[n-\epsilon,\,n+\epsilon]$ is in $\mathcal{O}(n)$.
So the complexity of the overall process is linear in the size of the
generated object.

An important point to mention is the case of polymorphic types. From a
theoretical point of view they fit in the framework. But from a
practical point of view it is hard to sample a ``polymorphic value''.
To deal with that limitation, the Boltzmann samplers are instantiated
only for concrete types \emph{e.g.} not for \lstinline|'a list| but for
\lstinline|int list|.

\section{Implementation and Experiments}\label{sec:implem}
We have implemented the work presented in the previous sections in  
a tool available at the url \url{https://github.com/ghilesZ/Testify}.
Our implementation relies on several state-of-the-art tools.  The
derivation of OCaml code from annotated OCaml source files is done
using the \emph{ppx} framework, as in \cite{Ya07,BaMa18}, which is a
form of generic programming~\cite{LaPe03}. Pre-processors using
\emph{ppx} are applied to source files before passing them on to the
compiler. They can be seen as self-maps over abstract syntax trees. In
our case, the source files are traversed to find OCaml type
declarations and derive their associated generators. These generators
are then used to provide inputs for the functions that must be tested.
We have implemented the techniques presented here for the global constraints that we have been able to 
identify in real data structures (BSTs, Sets, etc) namely \textit{alldiff}, \textit{increasing} and 
\textit{decreasing} (both strict and large versions). Note that to extend our implementation, \ie add a 
global constraint,, it is sufficient to add to the constraint solver a propagator for the said global 
constraint, as both the step of traversing the structure and the random generation procedure presented in section ~\ref{sec:solving} being common to all types.

The work done by Testify is divided into two phases: the first
is the pre-processing phase during which our tool collects some
information on the types needed to build the generators. The second is
the testing phase, where the generated code is executed to produce
inputs for the functions under test. Note that the pre-processing
phase is performed only once while the testing phase can be triggered
multiple times, each time one needs to run the tests.

We distinguish four kinds of types, for which we provide four
different synthesis techniques:

\begin{itemize}
\item For non recursive unconstrained
types (\eg \lstinline{int}, \lstinline{float * (int * int)} ... ) we determine at pre-processing time the function to be used as a generator. For that, we rely on the qcheck~\cite{qcheck} library, which provides the primitives for building and composing generators.
\item For non recursive constrained types (\eg
\lstinline{int[@satisfying fun x -> x >=0]}), we extract a
single CSP which is solved once, still at pre-processing
time. From this resolution is extracted a code that draws
uniformly solutions of this CSP and rebuild from them a value of the corresponding type. This is the method described in~\cite{ZDB21}.
\item For recursive unconstrained types (\eg lists, binary
trees), we build samplers by using the Arbogen~\cite{arbogen} tool. This tools implements the Boltzmann method presented in Sec.\ref{sec:boltzmann}.  The tuning of the Boltzmann parameter is done at pre-processing time while the shape generation, and the conversion of this shape to a value of the targeted type is done at testing time.
\item Finally, for recursive constrained types (\eg sorted lists,
  binary search trees), the previous techniques are mixed together to
  produce efficient generators: first, a targeted size $n$ is drawn,
  then, a shape of size $n$ is sampled. We then browse the generated
  shape, collecting constrained values to build a CSP as explained in
  Sec.\ref{sec:solving}. This CSP is then fed to the SICStus
  Prolog~\cite{sicstus} solver, which builds from it a generator using
  the \emph{PRT} library~\cite{Gotlieb10}. Finally we put together
  shapes and constrained values. All of these steps are made at
  testing time, that is every time we have to generate a value we must
  solve a CSP. This is arguably the bottleneck of our architecture,
  but experiments still demonstrate the usability of our method.
\end{itemize}

\subsection{Experiments}\label{sec:bench}
In this section, we focus on the performance of our automatically
derived generators. We measure the generation times (in seconds) obtained with our
method for different constrained recursive types and by varying the
size of the generated structure. The types we are interested in are:
lists sorted in ascending order, association lists with unique keys,
lists of pairs in ascending order ($(x,y) \leq (x', y) \Leftrightarrow
x \leq x' \wedge y \leq y'$), binary search trees (unbalanced),
functional maps (key-value stores as binary search trees) and
quadtrees. These types are among the most frequents in the literature,
and they only involve numerical constraints, which Testify is
able to manage.

\begin{figure}[ht]
\centering
\begin{tabular}{|c|p{4em}|p{6em}|p{6em}|p{6em}|}
\hline
  \textbf{Types} & \textbf{Targeted} & \textbf{Average} & \textbf{$\sharp$Objects} & \textbf{time}\\
  \hline
  
  \multirow{4}{*}{increasing_list}&
  10& 8.50& 2889& 0.020\\
  \cline{2-5}
  &100& 93.95& 13691& 0.004\\
  \cline{2-5}
  &1000& 948.57& 17763& 0.003\\
  \cline{2-5}
  &10000& 9392.45& 79& 0.757\\
  \hline
  
  \multirow{4}{*}{assoc_list}&
  10& 8.49& 2534& 0.023\\
  \cline{2-5}
  &100& 93.96& 11949& 0.005\\
  \cline{2-5}
  &1000& 947.87& 13660& 0.004\\
  \cline{2-5}
  &10000& 9406.92& 76& 0.786\\
  \hline
  
  \multirow{4}{*}{bicollect}
  &10& 6.99& 2492& 0.024\\
  \cline{2-5}
  &100 &93.04 &6418 &0.009\\
  \cline{2-5}
  &1000 &947.73 &16048 &0.003\\
  \cline{2-5}
  &10000 &9456.85 &1596 &0.037\\
  \hline
  
  \multirow{4}{*}{binary_tree}&
  10& 9.00& 238690& 0.001\\
  \cline{2-5}
  &100& 94.35& 21214& 0.001\\
  \cline{2-5}
  &1000& 948.00& 3416& 0.006\\
  \cline{2-5}
  &10000& 9740.00& 1500& 0.040\\
  \hline

  \multirow{4}{*}{map}
  &10 &9.00 &238690 &0.001\\
  \cline{2-5}
  &100 &94.37 &21208 &0.001\\
  \cline{2-5}
  &1000 &947.08 &3423 &0.006\\
  \cline{2-5}
  &10000 &9047.00 &1276 &0.047\\
  \hline

  \multirow{4}{*}{quad_tree}&
  10& 8.00& 3590507& 0.001\\
  \cline{2-5}
  &100& 93.88& 228357& 0.001\\
  \cline{2-5}
  &1000& 947.79& 23548& 0.002\\
  \cline{2-5}
  &10000& 9489.19& 2191& 0.027\\
  \hline
\end{tabular}
\caption{Generation time per object according to the size of the
  structure\label{fig:bench}}
\end{figure}

The experience was to sample as much as possible constrained
structures during one minute.  The results are shown in
Fig.\ref{fig:bench}.  For each type we report the size of the terms
(number of \lstinline|[@collect]| values) targeted, the average size
of the generated terms, the number of terms sampled and the average
time to sample one term.  The computer running the experiments has an
Intel Core i7-6700 CPU cadenced at 3.40GHz with 8 GB of RAM.

As expected, at least for the tree-like types, we observe that the
complexity is quite linear in the size of the sampled terms: the
Boltzmann method keeps its promises and the use of a constraint solver
proves to be fast enough to be used in our context.  For most of these
structures we manage to generate several hundred values per second, up
to a certain structure size. These results prove the relevance of our
method in the context of testing, as it can allow the user to
fine-tune the generators to decide whether he wants to test his
functions on several small structures and/or a few large ones.
However, we may note that sampling of lists is much slower than
sampling of trees. This is due to the fact that the Boltzmann method
is not tailored for regular languages (such as list). It would
probably be more efficient to use specialised algorithms for regular
languages such as the one of~\cite{BeGi12}.

\section{Related Work}\label{sec:related}
In this section we focus on related work dealing with constraint-based
generation techniques. Constraint-based generation of test data has
been exploited in white-box testing to produce inputs that will follow
some execution paths, as well as in functional testing to generate
constrained inputs.  In \cite{SenniF12}, Senni applies constraint
logic programming to systematically develop generators of structurally
complex test data, e.g. red-black trees, in the context of
Bounded-Exhaustive Testing.

PBT, as exemplified by Quickcheck for Haskell, has
been adapted to many programming languages but also to proof
assistants to test conjectures before proving them,
e.g. \cite{DHT2003,Bulwahn12,Paraskevopoulou15,Carlier10}. In
\cite{DHT2003} restricted classes of indexed families of types are
provided with surjective generators.  In \cite{Carlier10}, the authors
propose the FocalTest framework for testing - conditional -
conjectures about functional programs and for automatically generating
constrained values. In this work, CP global constraints are not used
and thus FocalTest does not take benefit from the corresponding
efficient filtering ad hoc procedures. \\
In the context of PBT of Erlang programs, De
Angelis \textit{et al} propose in \cite{Da&al2019} an approach to
automatically derive generators of values that satisfy a given
specification. Generation is performed via symbolic execution of the
specification using constraint logic programming.  A difference
between their approach and ours is that we craft a suitable
representation of a given type at static time, which is then
compiled into an efficient generator. In \cite{Da&al2019}, generators
are built at execution time, while testing, which ultimately leads to
a slower generation.\\
The Coq plugin QuickChick helps to test Coq conjectures as soon as
involved properties are executable. It allows the automatic synthesis
of random generators for algebraic data-types, recursive or not, and
also the definition of  simple inductive properties, e.g. a property
specifying binary search trees whose elements are between two bounds,
to be turned into random generators of constrained
values~\cite{LampropoulosPP18}.  
The approach is narrowing-based, like in~\cite{CDP2015}. Such a binary tree is built lazily while solving the constraints found in the inductive property while in Testify, the shape of the data structure is randomly chosen and then its elements are obtained by solving constraints.
This tool
comes with different primitives or mechanisms allowing for some
flexibility in the distribution of the sampled values. For example the
user can annotate the constructors of an inductive data-type with
weights that are used when automatically deriving generators. Furthermore, it also 
produces proofs of the generators correctness.
In \cite{CaDa09}, the authors adapt a Boltzmann model for random
generation of OCaml algebraic data-types, possibly recursive, but not
constrained. Generators are automatically derived from type
declarations. In \cite{CDP2015}, Claessen et al. propose an algorithm
that, from a data-type definition, a constraint defined as a Boolean
function and a test data size, produces random constrained values with
a uniform distribution. However the authors show that this uniformity
has a high cost. They combine this perfect generator with a more
efficient one based on backtracking. Limiting the class of constraints
and combining it with an efficient solving process, Testify can
generate constrained values with a uniform distribution in a
reasonable time.  Some work focus on the enumeration or sampling of
combinatorial structures, like lambda-terms, using Boltzmann samplers
\cite{Lescanne13}, Prolog mechanisms \cite{BodiniT17} or both
\cite{BendkowskiGT18}. These approaches are dedicated to objects of
recursive algebraic data-types with complex constraints, like typed
lambda-terms, closed lambda-terms, linear lambda-terms, etc. This kind
of constraints is out of reach of our tool whose objective is not only
to generate constrained values but also to provide the programmer with
syntactic facilities to specify them.

\section{Conclusion}\label{sec:conclusion}
We have proposed in this paper a technique based on declarative
programming, to derive generators of random and uniform values
for constrained recursive types.  We have proposed a small
description language for recursive structure traversal which
allows us to build a custom CSP for each term to be
generated. The code we generate is efficient, and outperforms a
naive generation technique based on rejection, and allows us to
generate large recursive structures quickly.  Starting from the
constraints attached to a type, we first sample the shape of the
value to generate and then build a CSP that encodes the valid
representations of the terms that have this shape. Then, our tool
uses the SICStus Prolog constraint solver to filter invalid
representations and produce a uniform solution sampler. Our
technique is integrated into the Testify framework, which embeds
these generators within a fully automatic test system. 
The generators derived by our framework are fast enough to allow
the user to run tests each time he compiles his code. This would
allow him to be able to detect bugs very quickly and fix them before
they become potentially harmful.
However, we still have a lot of work to do to improve Testify. For
example, we can extend the constraint language to be able to handle
types with shape constraints (\eg balanced trees). This would require adapting the Boltzmann technique to random sampling of tree
structures under constraints. Also, when dealing with a functional
language, functions as values cannot be avoided: it will be
necessary to have techniques for the derivation of generators for
functions, and explore what kind of constrained functions (monotonic,
bijective functions, etc.) appear in practice in programs. Moreover,
in this paper we have only studied \emph{tree-like} recursive
data-structures. Some structures do not fit into this
framework (\eg graphs, doubly linked lists) and it would be
interesting to see to what extent our methods adapt to these
structures.  Also, our current implementation tests functions by
generating any random input, disregarding their body. This is
naturally an important point of improvement. For example, one could
imagine a static analysis of the body of the function, to conduct the
input generation more precisely, and find bugs faster.  Finally, our
framework targets OCaml but the methods developed in this paper can be
adapted to most programming languages and proof assistants.


\bibliography{biblio}

\end{document}